# Ultrasensitive THz sensing with high-*Q* Fano resonances in metasurfaces

Ranjan Singh,[1,2,a),*] Wei Cao,[3,*] Ibraheem Al-Naib,[4,*] Longqing Cong[1,2], Withawat Withayachumnankul,[5] and Weili Zhang[3,b)]

[1]*Division of Physics and Applied Physics, School of Physical and Mathematical Sciences, Nanyang Technological University, Singapore 637371, Singapore*
[2]*Centre for Disruptive Photonic Technologies, School of Physical and Mathematical Sciences, Nanyang Technological University, Singapore 637371, Singapore*
[3]*School of Electrical and Computer Engineering, Oklahoma State University, Stillwater, Oklahoma 74078, USA*
[4]*Department of Physics, Engineering Physics and Astronomy, Queen's University, Kingston ON K7L 3N6, Canada*
[5]*School of Electrical & Electronic Engineering, The University of Adelaide, Adelaide, SA 5005, Australia*

(Dated: 27[th] June, 2014)

## Abstract

High quality factor resonances are extremely promising for designing ultra-sensitive refractive index label-free sensors since it allows intense interaction between electromagnetic waves and the analyte material. Metamaterial and plasmonic sensing has recently attracted a lot of attention due to subwavelength confinement of electromagnetic fields in the resonant structures. However, the excitation of high quality factor resonances in these systems has been a challenge. We excite an order of magnitude higher quality factor resonances in planar terahertz metamaterials that we exploit for ultrasensitive sensing. The low-loss quadrupole and Fano resonances with extremely narrow linewidths enable us to measure the minute spectral shift caused due to the smallest change in the refractive index of the surrounding media. We achieve sensitivity levels of $7.75 \times 10^3$ nm/ RIU with quadrupole and $5.7 \times 10^4$ nm/RIU with the Fano resonances which could be further enhanced by using thinner substrates. These findings would facilitate the design of ultrasensitive real time chemical and biomolecular sensors in the fingerprint region of the terahertz regime.

*These authors have equal contribution

[a)]Electronic Mail: ranjans@ntu.edu.sg

[b)]Electronic Mail: weili.zhang@okstate.edu

Recent years have seen tremendous progress in the field of plasmonics and metamaterials[1,2]. Metamaterials are artificially engineered medium designed from subwavelength building block resonators that are often referred to as "meta-atoms". The exotic properties of meta-atoms have led to very novel effects in controlling the electromagnetic fields that have given rise to negative refraction, super lenses, invisibility cloaks, slow light effects, and subwavelength resolution imaging[3,4]. Although significant progress has been made in demonstrating these functionalities with smart metamaterial device designs, there are still fundamental roadblocks to their high efficiency performance which has restricted their widespread practical implementation. Low performance of metamaterial and plasmonic devices is mainly due to the radiative and the non-radiative losses encountered in these systems [5]. Recently, several schemes have been attempted to minimize the losses such as by introducing gain, or using low loss dielectric materials, or by optimizing the shape and size of the metallic subwavelength structure[6-8].

Overcoming the losses would certainly enhance the performance of almost all metamaterial and plasmonic based devices. One of the significant device applications that plasmonic metamaterial has been proposed is that of a sensor at optical[9-12], infrared[13-17], and terahertz[18-25] frequencies. Ultrasensitive sensing performance could be only achieved by high quality ($Q$) factor resonances that have extremely narrow linewidths and are easy to detect. Metamaterial structures have a unique advantage of being able to support resonances at any desired frequency based on their structural size and these resonances are responsive to the changes in their effective index of refraction at their surface. As the $Q$ factor of the metamaterial increases, the photon lifetime in the resonator increases that leads to enhanced field interaction with the sample on the surface. Metamaterials have the potential to be a great sensing platform since they could be easily fabricated on different types of flexible and thin substrates [22,23]. However, the losses in metamaterials limit their sensing performance as the low quality factor resonances typically lacks the ability to detect small shifts in the resonance frequency due to the spectrally broad resonances in metamaterials[22-25]. Most of the previous work on metamaterial and plasmonic sensing depend on the low $Q$ resonances.

In this article, we have performed sensing using planar metasurfaces with an order of magnitude higher $Q$ factors than previously demonstrated. Low loss, high $Q$ quadrupole and Fano resonances are excited by breaking the symmetry of the metamaterial resonator structure, thus forming an asymmetric split ring resonator (ASR)[7,8,26,27]. The higher $Q$ resonances support strong interaction between the electromagnetic wave and the specific analyte. The sharp resonances of a low-loss high $Q$ metamaterial allows the detection of very small spectral shifts that occurs from the minute quantity of analyte interacting with the metamaterial resonator. We perform terahertz (THz) domain refractive index sensing with high $Q$ quadrupole resonance excited in one orientation and the high $Q$ Fano resonance is excited in the orthogonal orientation[8]. Terahertz sensing has received a lot of attention in recent times because of their significant scientific and technological potential in multidisciplinary fields [28,29]. Metasurfaces offer another important sensing platform at this frequency regime and there is a continued quest for enhancing the sensitivity of these planar devices.

We utilized a broadband terahertz time-domain spectrometer (THz-TDS) to characterize the sensing feature of high-$Q$ quadrupolar and Fano resonances. The photoconductive switch based spectrometer consists of four parabolic mirrors configured in an 8-F confocal geometry that provides excellent beam coupling between the transmitter and receiver[30,31]. The Gaussian beam of the THz pulse is focused to a frequency-independent beam waist of 3.5 mm for small sample characterization. The THz asymmetric split ring (TASR) metamaterial samples were fabricated on high-resistivity (4 kΩ·cm), double-side polished, 0.5-mm-thick n-type silicon substrate using photolithography and a 200-nm-thick aluminum was thermally metallized to form the TASRs. A 10-mm-thick, high-resistivity (4 kΩ·cm) silicon plate was placed in optical contact behind the metamaterial substrate to eliminate the Fabry Perot reflection from the back surface of the substrate that enabled a scan length of 200 ps in the time domain allowing frequency resolution of 5 GHz. Figure 1(a) shows the microscopic image of the TASR array and all the geometric dimensions of the structure. Asymmetry in the TASR is introduced by displacing the lower gap from the central vertical axis with $d$ =5 μm, where $d$ represents the lower gap displacement from the center. The sample array size is 10 mm × 10 mm. In the THz-TDS measurements, the metamaterial sample is placed midway between the transmitter and receiver in the far-field at the focused beam waist, and the transmitted terahertz pulses were measured at normal incidence such that the electric and magnetic fields of the incident terahertz radiation are in the plane of the metamaterial. Figure 1(b) shows the schematic of the quadrupole resonance current distribution for incident electric field along the *x*-axis and the corresponding excitation of a resonance with high quality factor of 65 is shown in Fig. 1(d). A different mode with anti-parallel current distribution, as shown in Fig. 1(c), typically known as the Fano resonance is excited with the incident electric field pointing towards the *y*-axis and the asymmetric line shaped resonance with $Q$ factor of 28 is excited. Therefore, the planar TASR metamaterial used in this work is capable of supporting high $Q$ resonances for electric field in orthogonal directions enabling ultrasensitive sensing when an analyte layer is deposited on top of the TASR planar metamaterial surface, as shown in Figs. 1(b) and 1(c). The corresponding red shift of quadrupole and Fano resonance frequencies could be clearly seen in Figs. 1(d) and 1(e).

Different thicknesses of photoresist are spun coated on the planar TASR samples. Figure 2(a) shows the simulated transmission response of the metamaterial where we observe gradual red shifting of the quadrupole resonance with increasing thickness of the spin coated photoresist overlayer. The red shift of quadrupole resonance for film thicknesses 1, 4, 8, and 16 µm are 8.38, 14.53, 17.41, and 19.5 GHz, respectively when compared with the quadrupole resonance frequency without any analyte on top of the meta-surface, i.e. 1.13 THz. Figure 2(b) shows the measured data which agree well with the simulations. The refractive index of the spin coated photoresist is $n = 1.6$. The largest amount of red shift occurs for the 1 µm thick photoresist. The inset of Fig. 2(b) shows current distribution of the quadrupole resonance mode where the pairs of diagonally anti parallel currents ensure extremely low radiation losses. We would later discuss the electric field distribution of the quadrupole mode and observe the strong fields excited at this resonance in the TASR structure. The red shift is mainly caused by the increase in the capacitance of the TASR gaps. The photoresist analyte fills the gaps which also leads to the enhancement of electric field in the TASR gaps.

The metamaterial samples were also measured for the orthogonal direction excitation at which asymmetric line shaped Fano resonance is excited at 0.52 THz without any photoresist coating. For overlayer thicknesses of 1, 4, 8, and 16 µm, the corresponding red shift of the Fano resonance is 10, 21.3, 26.1, and 29 GHz, respectively, as shown in Figs. 3(a), which represents the simulated transmission. Figure 3(b) is the measured transmission and good agreement is observed between the data and simulation. The largest degree of redshift in the Fano resonance frequency was once again for the first overlayer coating of 1 µm. The total shift of 29 GHz for 16 µm thickness analyte is significantly higher than the total shift (19.5 GHz) in quadrupole resonance. This highlights the higher sensitivity of the Fano resonance over the quadrupolar resonance which we would discuss in detail later. The inset of Fig. 3(b) shows the antiparallel current directions in the TASR at the Fano resonance frequency. The antiparallel current distribution is key in reducing the radiation losses in the structure and it also facilitates a very weak coupling with the free space.

In order to understand the redshift of the quadrupole and the Fano resonances, we used a commercial Maxwell equation solver CST Microwave Studio to investigate the electric field at the resonances with and without the analyte overlayer on top of the metasurface. Column 1 in Fig. 4 reveals the field distribution at various resonances without any analyte and column 2 shows the field at the corresponding shifted resonances with 16 µm photoresist overlayer. Figures 4(a), 4(c), and 4(e) show the electric field in TASR at different frequencies when the incident terahertz field is along the $x$ − axis. In Fig. 4(a), we see the fields at the quadrupole resonance peak at 1.13 THz. A closer observation clearly reveals the four nodes in the field distribution due to which we define the resonance as the quadrupolar mode. In the adjacent Fig. 4(b) the electric fields in the two split gaps become stronger compared to that in Fig. 4(a) due to the spin coated photoresist 16 µm dielectric layer. The photoresist supports the field enhancement as it occupies the volume in the split gap on top as well as bottom arm of TASR that act as dual capacitors in each unit cell. Figure 4(c) shows the electric field at the first resonance dip (at 1.096 THz) for the $x$ axis terahertz excitation and we can notice that the shorter left side arm of the TASR is excited with two distinct electric field nodes, thus acting as a dipolar resonant excitation. Similarly, in Fig. 4(e), the simulated electric field is at the second resonance dip (1.158 THz) at which only the right hand side longer wire is excited with two observable nodes. Both of these resonance dips are dipolar excitations individually in each of the metallic wire that forms the TASR and the corresponding red shifted resonant electric fields with 16 µm overlayer has been shown in Figs. 4(d) and 4(f), respectively. With the analyte, we observe field enhancement at the edges of each metallic wire for the dipolar excitations.

After probing the two resonance dips and the single peak of the quadrupole resonance, we altered the terahertz excitation (with E along the *y* axis) and investigated the electric field at the dip of the Fano resonance at 0.515 THz, as shown in Fig. 4(g). With a closer look at the fields, it appears that at the Fano resonance their intensity in the split gaps of TASR is much stronger compared to that of the quadrupolar resonance. This observation is key in understanding that these tightly confined fields at the Fano resonance excitation are extremely sensitive to its surroundings, thus we get almost twice as much red shift of the resonance than that compared to the quadrupolar resonances. As it could be seen in Fig. 4(h), the field confinement is further enhanced with coating a dielectric analyte on top of the metasurface.

Since the change in the resonance frequency of quadrupole and the Fano resonances were different for the same thickness of spin coated analyte on the metasurface, we analyzed the sensitivity of both resonances by maintaining constant thickness of the analyte and varying the refractive index of the analyte to be sensed. Figure 5(a) shows the shift in the quadrupole resonance with a constant thickness of 4 μm but different refractive index analyte. The total shift in the quadrupole resonance frequency by changing the refractive index of analyte from $n = 1$ to $n = 1.6$ is found to be about 15 GHz. We repeated the analysis for the Fano resonance shown in Fig. 5(b) and the total shift in this case is observed to be 22 GHz. We plotted the resonance shift versus the change in refractive index of the analyte for a constant thickness of 4 μm in Fig. 5(c) and estimated the sensitivity of both resonances. The quadrupole and Fano resonance sensitivities turned out to be 23.9 and 36.7 GHz/refractive index unit (RIU), respectively. We converted these numbers into Δλ/RIU by using $\left|\frac{d\lambda}{dn}\right| = \frac{c}{f_0^2} * \frac{df}{dn}$, where '*c*' is the speed of light, $f_0$ is resonance frequency, and $n$ represents the refractive index of the analyte. In terms of $\Delta\lambda$/RIU, the sensitivities that we obtained for the 4 μm thick analyte with quadrupole and Fano resonances are 5.62 x $10^3$ nm/RIU and 4.23 x $10^4$ nm/RIU, respectively. The sensitivity values are higher than previously reported sensitivities using planar metamaterial structures on identical thickness substrates [24, 25].

We further investigated the analyte thickness dependent sensitivity of the TASR metasurface through rigorous simulations as shown in Fig. 6 (a). We observe that the sensitivities of both of the resonances increase exponentially with the increasing analyte thickness and eventually saturate at about 16 μm thickness of the analyte overlayer. We also notice that the Fano resonance has significantly higher sensitivity than the quadrupole which saturates at lower maximum sensitivity value. The maximum sensitivities are found to be 49.3 GHz/ RIU ($5.7\times10^4$ nm/RIU) for the Fano resonance and 33 GHz/ RIU ($7.75\times10^3$ nm/ RIU) for the quadrupole resonance. The difference in maximum sensitivities of the two resonances can be attributed to the different mode profiles of the resonances. An easier route to further enhance the sensitivity of the TASR metasurfaces could be the use of ultrathin substrates since it allows stronger interaction of analyte with the intense field in the capacitive gaps that otherwise remains mostly confined in thick substrates. Thus, we probed the frequency shift of both resonances by reducing the thickness of the silicon substrate. The analyte coated on the metasurface has a fixed thickness of 1 μm with $n = 1.6$ (same as the photoresist index). The metasurface becomes extremely sensitive and shows much larger frequency shift below substrate thickness of 20 μm as shown in Fig. 6(b). This occurs due to the strong interaction of analyte and the enhanced electric field in the capacitive gaps of the metasurface. Large resonance frequency shift shows that thinner substrates can be applied to further enhance the sensitivities of the high $Q$ metasurfaces.

In summary, we have demonstrated high $Q$ ultrasensitive sensing via strong light-matter interaction at the ultra-sharp quadrupole and Fano resonances in the fingerprint terahertz spectral domain. High resolution terahertz measurements allowed us to measure the minute shift in the resonance frequencies as thin layers of analyte bonded with the metasurface. The sensitivities achieved by using the high $Q$ resonant metasurfaces in this work outperform other schemes based on low $Q$ planar metamaterial and plasmonic structures. We believe that the impact of our results lies in identifying high $Q$ terahertz metasurface based ultrasensitive sensing that would open up avenues for advanced design and realization of real time on-chip chemical and biomolecular detection with spectral signatures in the terahertz regime.

**Figure Captions**

**FIG. 1. High $Q$ planar meta-sensor.** (a) Microscopic image of the terahertz asymmetric split-ring (TASR) metamaterial array with the detailed geometric dimensions (b) TASR unit cell where the quadrupole resonance is excited and the analyte photoresist layer is deposited on top of the metasurface (c) the transmission spectra of the quadrupole resonance with and without the analyte layer (d) TASR unit cell where the Fano resonance is being excited (e) transmission spectra of the Fano resonance with and without the analyte layer.

**FIG. 2. Shift in Quadrupole resonance.** (a) Calculated and (b) measured transmission spectra with different thickness of photoresist coating on the metasurface. Quadrupole resonance is excited when E field is along the *x* axis.

**Fig. 3. Shift in the Fano resonance.** (a) Calculated and (b) measured transmission spectra with different thickness of photoresist coating on the metasurface. Fano resonance is excited when E field is along the y axis.

**Fig. 4. Electric field distribution at Quadrupole and Fano resonance with and without analyte binding.** Simulated electric field distribution at (a), (b) 1.13 THz, quadrupole resonance peak, (c), (d) 1.096 THz, dipole resonance dip, (e), (f) 1. 158 THz, dipole resonance dip, (g), (h) 0.515 THz, Fano resonance dip with 0 μm and 16 μm photoresist overlayer thickness coated on the metasurface.

**Fig. 5**. **Sensitivities of Quadrupole and Fano resonances**. Simulated shift in (a) quadrupole resonance and (b) Fano resonances when 4 μm constant thickness analyte with different refractive indices are coated on the metasurface. (c) Quadrupole and Fano resonance shifts with the change in refractive indices.

**Fig. 6. Sensitivity of different analyte thicknesses and enhanced resonance shift in thin substrates**. Simulated (a) Sensitivities of different thickness analyte overlayer (b) Resonance frequency shift when 1 μm thick analyte with refractive index $n = 1.6$ is coated on metasurfaces with decreasing substrate thicknesses.

**References**

1	Zheludev, N. I. & Kivshar, Y. S. From metamaterials to metadevices. *Nature materials* **11**, 917-924 (2012).
2	Liu, Y. & Zhang, X. Metamaterials: a new frontier of science and technology. *Chemical Society Reviews* **40**, 2494-2507 (2011).
3	Shalaev, V. M. Optical negative-index metamaterials. *Nature photonics* **1**, 41-48 (2007).
4	Soukoulis, C. M., Kafesaki, M. & Economou, E. N. Negative-Index Materials: New Frontiers in Optics. *Advanced materials* **18**, 1941-1952 (2006).
5	Boltasseva, A. & Atwater, H. A. Low-loss plasmonic metamaterials. *Science* **331**, 290-291 (2011).
6	Xiao, S. *et al.* Loss-free and active optical negative-index metamaterials. *Nature* **466**, 735-738 (2010).
7	Fedotov, V., Rose, M., Prosvirnin, S., Papasimakis, N. & Zheludev, N. Sharp trapped-mode resonances in planar metamaterials with a broken structural symmetry. *Physical review letters* **99**, 147401 (2007).
8	Cao, W. *et al.* Low-loss ultra-high-Q dark mode plasmonic Fano metamaterials. *Optics letters* **37**, 3366-3368 (2012).
9	Anker, J. N. *et al.* Biosensing with plasmonic nanosensors. *Nature materials* **7**, 442-453 (2008).
10	Lahiri, B., Khokhar, A. Z., De La Rue, R. M., McMeekin, S. G. & Johnson, N. P. Asymmetric split ring resonators for optical sensing of organic materials. *Optics express* **17**, 1107-1115 (2009).
11	Liu, N., Tang, M. L., Hentschel, M., Giessen, H. & Alivisatos, A. P. Nanoantenna-enhanced gas sensing in a single tailored nanofocus. *Nature materials* **10**, 631-636 (2011).
12	Papasimakis, N. *et al.* Graphene in a photonic metamaterial. *Optics express* **18**, 8353-8359 (2010).
13	Cubukcu, E., Zhang, S., Park, Y.-S., Bartal, G. & Zhang, X. Split ring resonator sensors for infrared detection of single molecular monolayers. *Applied Physics Letters* **95**, 043113 (2009).
14	Liu, N. *et al.* Planar metamaterial analogue of electromagnetically induced transparency for plasmonic sensing. *Nano letters* **10**, 1103-1107 (2009).
15	Wu, C. *et al.* Fano-resonant asymmetric metamaterials for ultrasensitive spectroscopy and identification of molecular monolayers. *Nature materials* **11**, 69-75 (2012).
16	Zhao, J., Zhang, C., Braun, P. V. & Giessen, H. Large-Area Low-Cost Plasmonic Nanostructures in the NIR for Fano Resonant Sensing. *Advanced Materials* **24**, OP247-OP252 (2012).
17	Yang, Y., Kravchenko, I. I., Briggs, D. P. & Valentine, J. High Quality Factor Fano-Resonant All-Dielectric Metamaterials. *arXiv preprint arXiv:1405.3901* (2014).
18	Al-Naib, I. A. I., Jansen, C. & Koch, M. Thin-film sensing with planar asymmetric metamaterial resonators. *Applied Physics Letters* **93**, 083507 (2008).
19	Debus, C. & Bolivar, P. H. Frequency selective surfaces for high sensitivity terahertz sensing. *Applied Physics Letters* **91**, 184102 (2007).
20	Driscoll, T. *et al.* Tuned permeability in terahertz split-ring resonators for devices and sensors. *Applied Physics Letters* **91**, 062511 (2007).

21  Ng, B. *et al.* Lattice resonances in antenna arrays for liquid sensing in the terahertz regime. *Optics express* **19**, 14653-14661 (2011).
22  Tao, H. *et al.* Metamaterials on paper as a sensing platform. *Advanced Materials* **23**, 3197-3201 (2011).
23  Tao, H. *et al.* Performance enhancement of terahertz metamaterials on ultrathin substrates for sensing applications. *Applied Physics Letters* **97**, 261909 (2010).
24  Withayachumnankul, W. *et al.* Sub-diffraction thin-film sensing with planar terahertz metamaterials. *Optics express* **20**, 3345-3352 (2012).
25  O'Hara, J. F. *et al.* Thin-film sensing with planar terahertz metamaterials: sensitivity and limitations. *Optics Express* **16**, 1786-1795 (2008).
26  Luk'yanchuk, B. *et al.* The Fano resonance in plasmonic nanostructures and metamaterials. *Nature materials* **9**, 707-715 (2010).
27  Miroshnichenko, A. E., Flach, S. & Kivshar, Y. S. Fano resonances in nanoscale structures. *Reviews of Modern Physics* **82**, 2257 (2010).
28  Tonouchi, M. Cutting-edge terahertz technology. *Nature photonics* **1**, 97-105 (2007).
29  Ferguson, B. & Zhang, X.-C. Materials for terahertz science and technology. *Nature materials* **1**, 26-33 (2002).
30  Grischkowsky, D., Keiding, S., Exter, M. v. & Fattinger, C. Far-infrared time-domain spectroscopy with terahertz beams of dielectrics and semiconductors. *JOSA B* **7**, 2006-2015 (1990).
31  Azad, A. K., Dai, J. & Zhang, W. Transmission properties of terahertz pulses through subwavelength double split-ring resonators. *Optics letters* **31**, 634-636 (2006).

**FIG. 1**

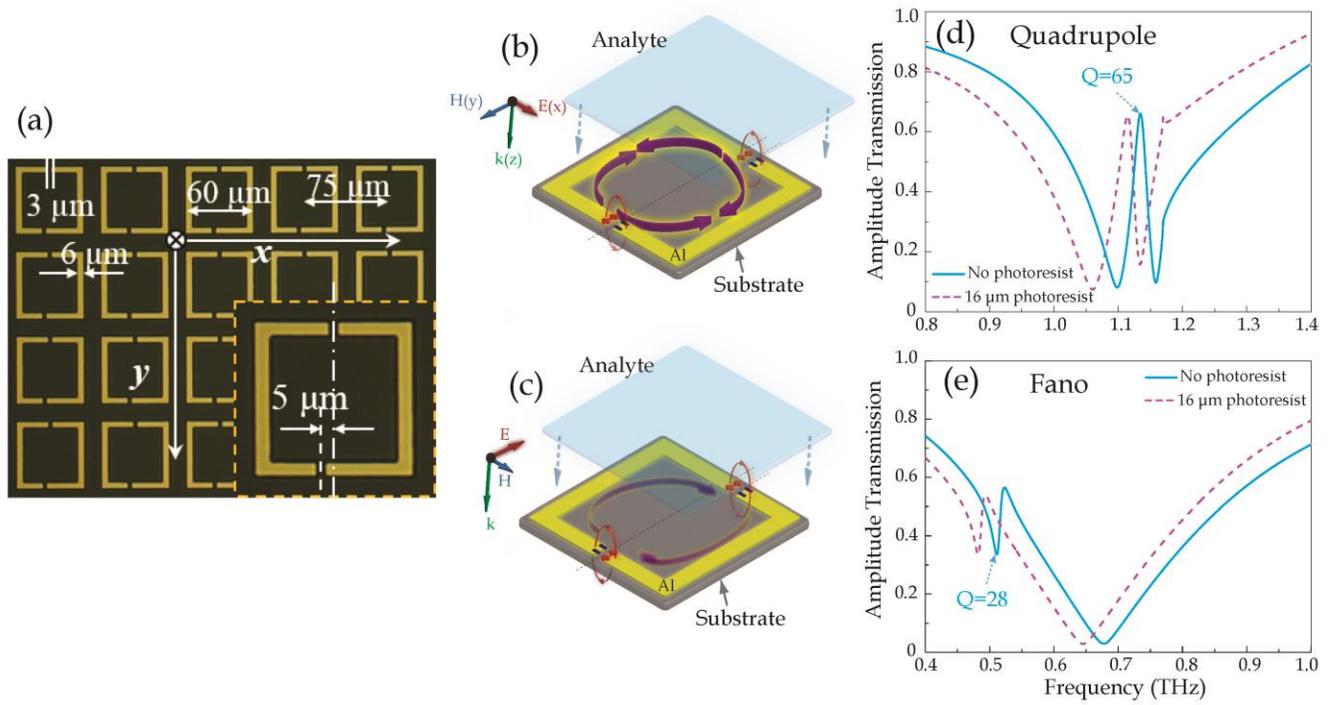

**FIG. 2**

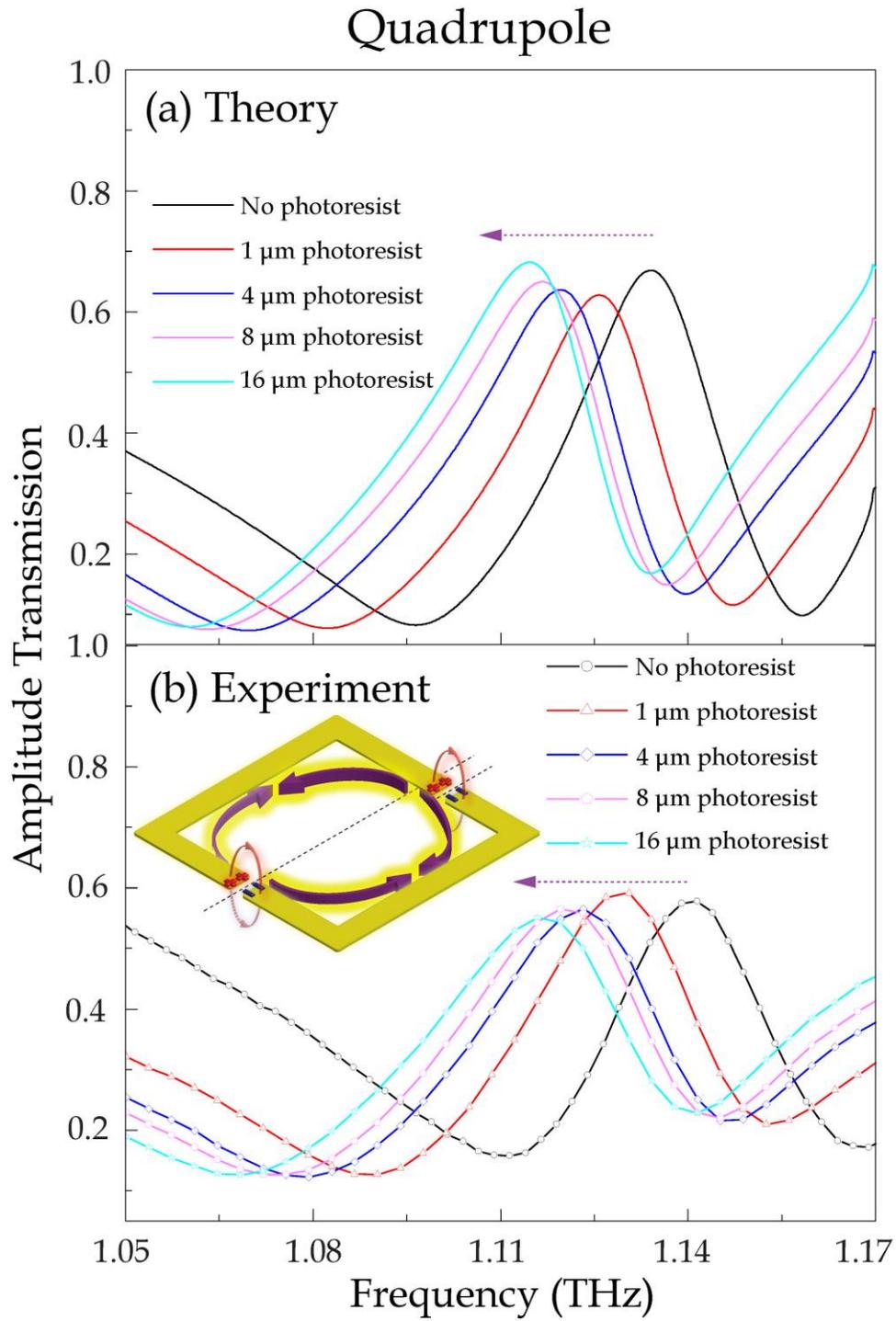

**FIG. 3**

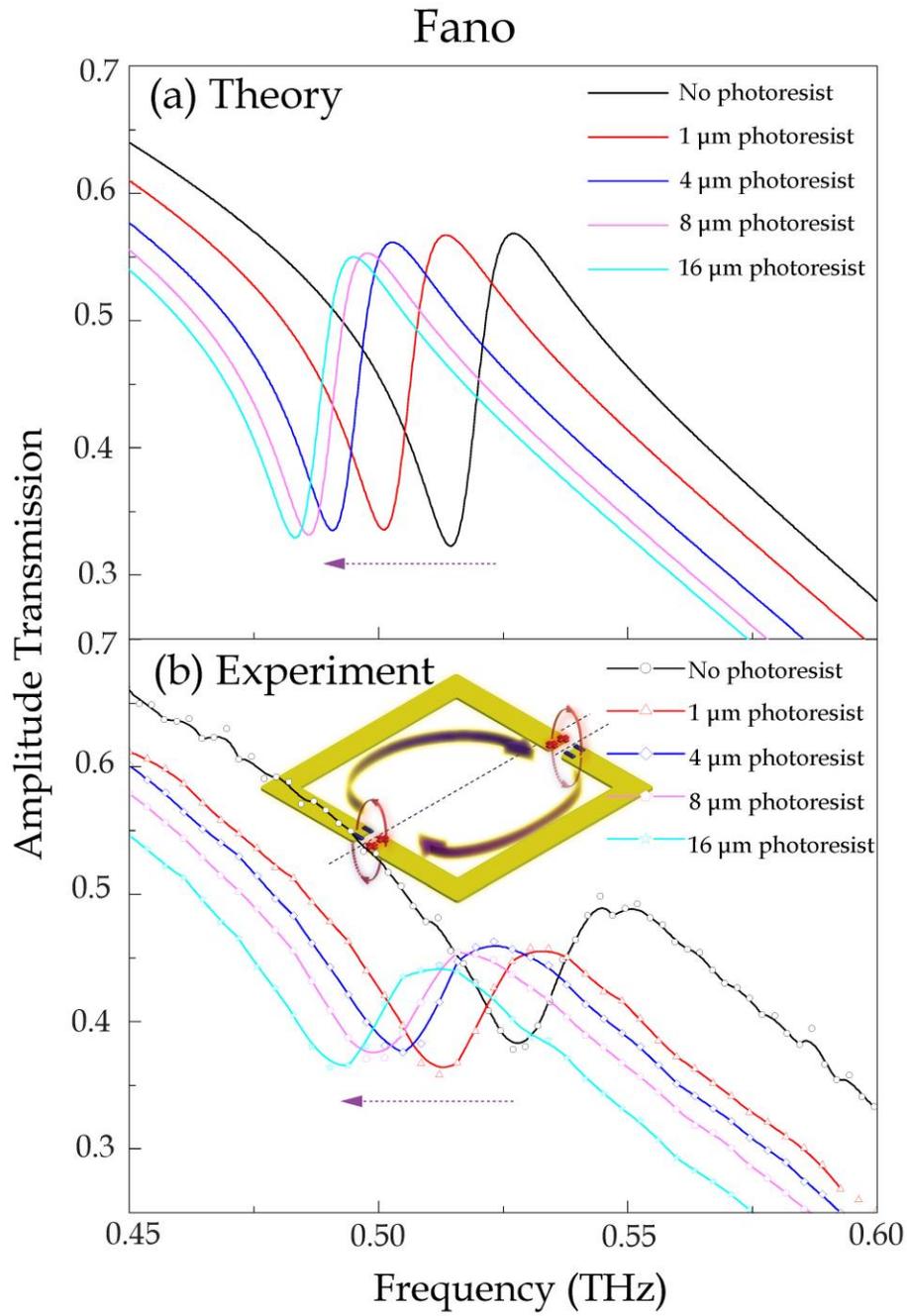

**FIG. 4**

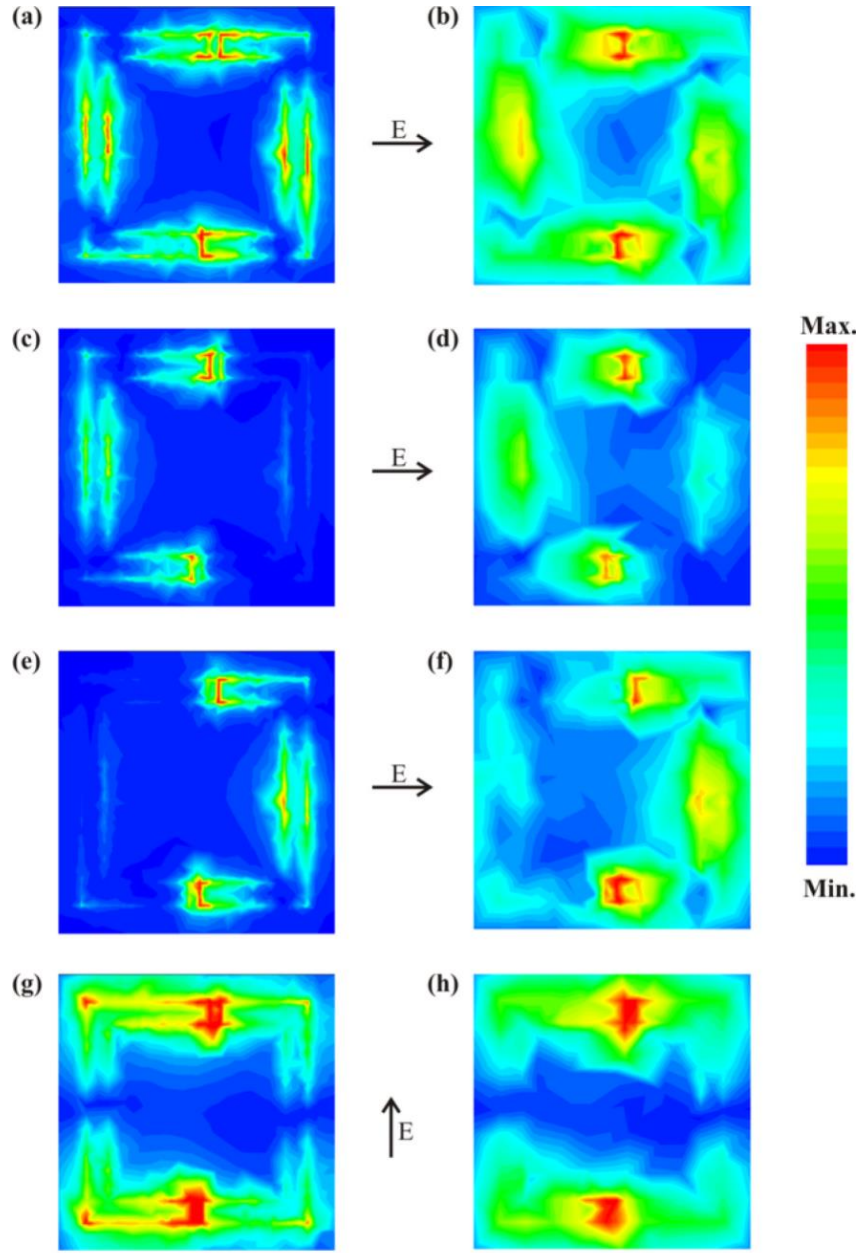

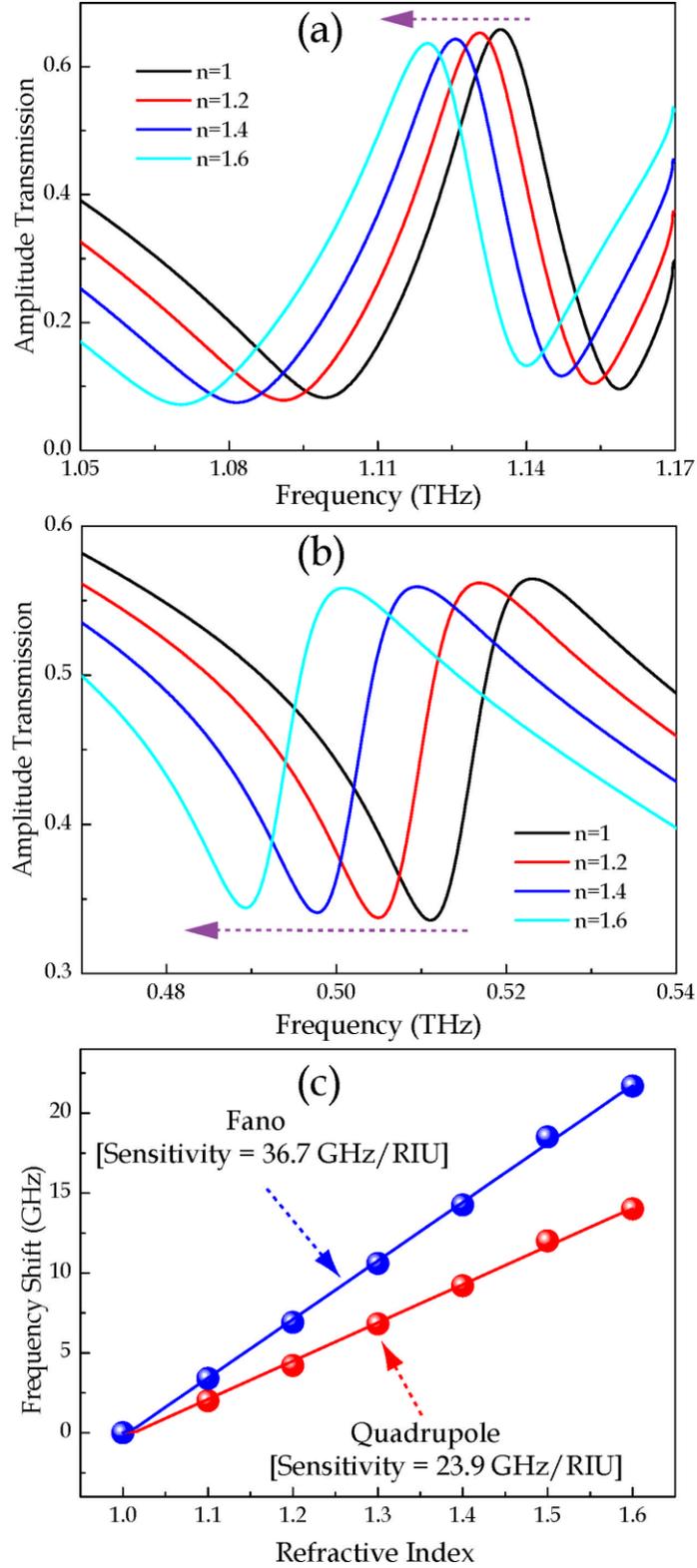

**FIG. 6**

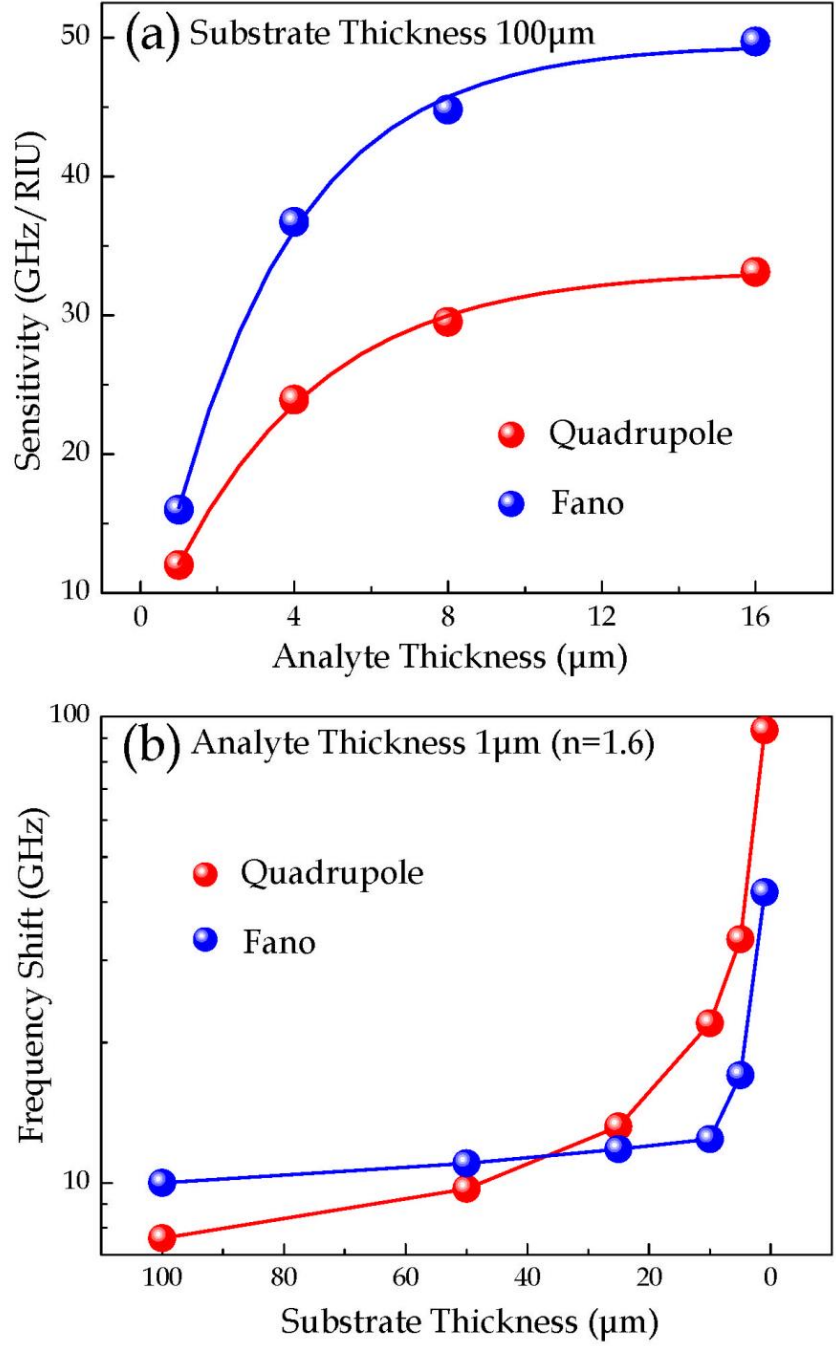